# Acoustic-based Object Detection for Pedestrian Using Smartphone

ZI WANG, Florida State University, USA
SHENG TAN, Florida State University, USA
LINGHAN ZHANG, Florida State University, USA
JIE YANG, Florida State University, USA

Walking while using a smartphone is becoming a major pedestrian safety concern as people may unknowingly bump into various obstacles that could lead to severe injuries. In this paper, we propose ObstacleWatch, an acoustic-based obstacle collision detection system to improve the safety of pedestrians who are engaged in smartphone usage while walking. ObstacleWatch leverages the advanced audio hardware of the smartphone to sense the surrounding obstacles and infers fine-grained information about the frontal obstacle for collision detection. In particular, our system emits well-designed inaudible beep signals from the smartphone built-in speaker and listens to the reflections with the stereo recording of the smartphone. By analyzing the reflected signals received at two microphones, ObstacleWatch is able to extract fine-grained information of the frontal obstacle including the distance, angle and size for detecting the possible collisions and to alert users. Our experimental evaluation under two real-world environments with different types of phones and obstacles shows that ObstacleWatch achieves over 92% accuracy in predicting obstacle collisions with distance estimation errors at about 2 cm. Results also show that ObstacleWatch is robust to different sizes of objects and is compatible to different phone models with low energy consumption.

CCS Concepts: • **Human-centered computing** → **Ubiquitous and mobile computing systems and tools**;

Additional Key Words and Phrases: Mobile Sensing, Pedestrian Safety, Acoustic Ranging, Acoustic Localization, Smartphone

## 1 INTRODUCTION

Due to the increasing popularity of mobile devices in recent years, pedestrians nowadays are more likely to engage in smartphone related activities while walking, such as texting, watching video, surfing the Internet, and playing games. This is known as distracted walking [32, 35], which could lead to severe injuries. For example, distracted pedestrians may unknowingly bump into various objects along the walking path such as trees, trash bin, and other pedestrian, resulting in various injuries like sprains, concussions, fracture and even death. Existing studies show pedestrian deaths have increased since 2009 and the distracted walking is becoming a major pedestrian

Authors' addresses: Zi Wang, Florida State University, 1017 Academic Way, Tallahassee, FL, 32306, USA, ziwang@cs.fsu.edu; Sheng Tan, Florida State University, 1017 Academic Way, Tallahassee, FL, 32306, USA, tan@cs.fsu.edu; Linghan Zhang, Florida State University, 1017 Academic Way, Tallahassee, FL, 32306, USA, lzhang@cs.fsu.edu; Jie Yang, Florida State University, 1017 Academic Way, Tallahassee, FL, 32306, USA, jie.yang@cs.fsu.edu.

safety concern [2, 23]. Indeed, the percentage of pedestrians' injuries related to mobile phone distraction has increased from 0.58% in 2004 to 3.67% in 2010 [27]. This trend is likely to continue due to the increasing number of distracted smartphone users. In addition, there is an increasing number of blind and visually impaired people who face challenges in interacting with surroundings, especially when walking on the street with obstacles.

To mitigate the risk of smartphone related pedestrian injuries, various infrastructure supports and policies have been proposed. For example, in Washington D.C., a special "cellphone lane" has been set up to separate pedestrians who use smartphones while walking from other pedestrians. In Delaware, decals have been installed to prompt pedestrians to look up at crosswalks and sidewalks of certain busy intersections. Moreover, numerous countries or states have attempted to pass bills to reduce the accident rate by imposing fines over smartphone related distracted walking behaviors [23]. On the other hand, the automobile companies are increasingly deploying various sensors on their vehicles for detecting distracted pedestrians [1, 9, 36].

There are also active research efforts in developing mobile sensing systems to alert distracted pedestrians with either smartphone built-in or dedicated sensors proposed in recent years. For example, Walksafe [40] utilizes smartphone front camera to detect the incoming cars when the user is talking on the phone. This system, however, is not able to detect the surrounding obstacles on the road. Lookup proposed by S. Jain *et al.* exploits motion sensors installed on shoes to identify losing steps when user walking through transitions between path and road [11]. Such an approach requires additional shoes mounted sensors and cannot detect obstacle collisions. Auto++ [17] utilizes smartphone to passively record acoustic sounds emitted from nearby vehicles to identify the direction and number of moving cars. This work, however, works only for running vehicles that actively emit acoustic signals. More recently, Tung *et al.* proposed a smartphone-based system, BumpAlert [37], which senses the nearby objects with the camera as well as the built-in speaker and microphone. BumpAlert however only provides coarse-grained information about the obstacle (i.e., only the distance), which is insufficient to detect the likelihood of colliding with the obstacle, triggering false alerts that distract users. Although these approaches improve pedestrian safety in various aspects, they either require additional sensors or can not provide sufficient information for fine-grained obstacle collision detection.

In this paper, we introduce ObstacleWatch, a smartphone-based obstacle collision detection system to improve the safety of pedestrians who are engaged in smartphone usage while walking. The system can also be used as visual assistant to help blind and visually impaired people to avoid obstacle collisions. With only the smartphone built-in speaker and microphones, ObstacleWatch can effectively detect collisions with the frontal obstacle when people are using (holding) smartphones while walking. ObstacleWatch achieves accurate collision detection by sensing the fine-grained information of the frontal obstacle including the size of the obstacle, the distance and the angle between the walking user and the obstacle.

In particular, our system exploits an acoustic sensing approach that takes advantages of the audio hardware advancements on smartphones for extracting the fine-grained information of the obstacles. Although nowadays the high definition audio capabilities supported by smartphones target at audiophiles, such capabilities can be leveraged to sense the surrounding obstacles with a wider frequency range and a higher sampling frequency at both the top and the bottom microphones. For example, current smartphones (e.g. Galaxy S5, S8+, and iPhone 7 and 8) have good frequency responses up to 23kHz. Such a high frequency response range has the advantages of inaudible to human ears and distinguishable from environmental noises. In addition, current smartphones' OSs and audio chips support playback and recording at 192kHz sampling frequency, which enables us to accurately pinpoint the obstacles at sub-centimeter level resolution. Moreover, the dual-microphone equipped on virtually all smartphones can be leveraged to estimate the relative position between the user and the frontal obstacle for collision detection.

Our system thus re-uses the smartphone as an acoustic sonar, which emits a well-designed beep signal at the inaudible frequency band from the build-in speaker and records the signal reflections from the obstacles at both the top and bottom microphones. Based on the signal reflections, we first locate the closest frontal obstacle and extract its fine-grained information. The distance between the obstacle and the user can be estimated based on the round trip time of the reflected signals. Then, the relative angle between the walking trajectory and the obstacle is sensed by leveraging the time-difference-of-arrival (TDoA) of the reflected signals to the two microphones (i.e., the top and bottom microphones) of the smartphone. To obtain an accurate TDoA, we develop an scheme to mitigate the impact of the human body reflections to improve the accuracy of TDoA estimation. Moreover, we leverage the fact that the size of the object has direct impact on the signal reflections to infer the object size. By combining these fine-grained information of the frontal obstacle, our system predicts the likelihood of the obstacle collision and alerts the user accordingly.

To evaluate the performance of our system, we conduct experiments in both a daily route near supermarket which represents a relatively open space with multiple types of obstacle, and a daily route of students in a campus that represents a crowded environment. We evaluate our system using three different types of smartphones under various experimental settings. Experimental results show that our system is highly effective in detecting obstacle collisions. The contributions of our work are summarized as follows.

- We develop ObstacleWatch, a smartphone-based obstacle collision detection system to improve the safety of distracted pedestrians with only the smartphone built-in speaker and microphones.
- Our system is the first to extract fine-grained information of the obstacle for collision detection by taking advantages of the dual-microphone and the advanced audio hardware on current smartphones. We also propose a scheme to mitigate the impact of the body reflections to the accuracy of the distance estimation between the obstacle and the two microphones of the phone.
- Our extensive experimental results show that ObstacleWatch achieves over 92% collision detection accuracy with about 2 cm distance estimation errors. Results also show that ObstacleWatch is able to work with different phone models under different real-world environments.

## 2 RELATED WORKS

In this section, we discuss the research works related to improve walking safety of smartphone users and acoustic localization. Existing pedestrian safety systems can be divided into two categories: non-acoustic-based systems and acoustic-based systems.

**Non-acoustic based Pedestrian Safety System.** This body of work focuses on sensing surrounding environments of the pedestrians and provide them with active feedback using motion sensors, cameras, and infrared [8, 47]. For example, Lookup [11] utilizes shoe-mounted motion sensors to identify the scenario when a user steps in an intersection based on surface gradient and step motion changes. Walksafe [40] uses smartphone camera to capture the images of incoming cars and feed into an image classifier to avoid car accident. Tang *et al.* exploit a vision-based approach with the back camera of the smartphone to identify the tactile paving deployed on the street and reduce the risks of distracted walking [34]. TerraFirma [12] introduced by Jain *et al.* also utilizes the smartphone camera to recognize the material and texture of ground surface to detect if user is walking in safe locations. pSafety [18] developed by Lin *et al.* alert users of potential collisions by detecting if the predicted area of pedestrians and vehicles are overlapped based on the GPS information. Ishikawa *et al.* propose a system to detect road anomaly and recognize avoidance behaviors of pedestrians by measuring azimuth changes from motion sensors as well as GPS information of the smartphone [10]. Vinayaga *et al.* design an activity recognition framework to detect pedestrian distraction using motion sensors such as accelerometers and gyroscopes on wearable devices [38]. A crowd sensing system to tell whether trail segments are risky to climb by KIM *et al.* also makes use of data from accelerometer on smartphones [14]. Liu *et al.* developed a infrared based system InfraSee [20], which is able to detect sudden change of ground by measuring the distance of the ground surface to the infrared sensor on smartphone. Moreover, Riaz *et al.* proposed a system SightSafety [30], which is a hybrid system that combines both wireless sensor network and GPS information to avoid vehicle/pedestrian collisions.

David *et al.* make a series of studies and design radio based collision avoidance systems to enhance the safety of the pedestrian [4, 7, 39]. Although the above mentioned research efforts focus on various pedestrian safety scenarios, none of them consider the possible collisions with obstacles along the walking road.

**Acoustic-based Pedestrian Safety System.** This category of work aims to sense the surrounding environment with acoustic signals. For instance, UltraSee [41] is able to detect the sudden changes of the ground when a user steps off the sidewalk. This is done by augmenting smartphone with ultrasonic sensors to detect abrupt distance change between the phone and the ground. Li *et al.* propose an acoustic based system Auto++ [17], which processes the acoustic signal captured by smartphone microphones to detect approaching cars from different directions. Auto++ passively records environmental sound and uses machine learning technique to extract features such as TRF to identify the direction and numbers of approaching cars. It work only for the moving cars who actively emit acoustic noises. However, in our scenario the static obstacles do not emit any acoustic sound. More recently, Tung *et al.* propose a smartphone based solution BumpAlert [37], which estimates the distance from the user to the nearby objects with smartphone camera, speaker and microphones. BumpAlert however only provides the distance information, which is insufficient to make accurate collision detection. Moreover, BumpAlert requires to use camera to further validate the obstacle in front of a user, which imposes additional energy consumption and may affect the user experience.

**Acoustic Localization Sensing.** Acoustic localization systems utilize sound to determine the distance and direction of the sound source or the reflectors. Such process can be done either actively or passively. Active localization can be achieved by combining the acoustic signal measurements at various receivers and the geometric relationships between receivers [46]. Early work proposed by Schau *et al.* uses an alternative intersecting spheres method for passive acoustic localization utilizing TDoA among different receivers [31]. Mohan *et al.* developed a system consists of sensor arrays to derive location of multiple acoustic sources using a coherence test [22]. System proposed by Liu *et al.* enables keystroke snooping by localizing the typing sound source, it obtains TDoA measurements by calculating the cross correlation of received signal in time domain [19]. A driver detection system presented by Yang *et al.* distinguishes a driver and a passenger by leveraging car speakers and mobile microphones [44]. The work proposed by Zhang *et al.* demonstrated internal phase difference (IPD) can also be utilized to find the TDoA [45]. In order for such approach to work, a high signal to noise rate is required which is difficult to acquire in distracted walking scenarios. Acoustic-based ranging system BeepBeep introduced by Peng *et al.* estimate the distance between smartphones and by counting samples difference between beeps each device can get two-way time of flight of the beeps and solve the challenge of synchronization [29].

There are also other related works that use acoustic signal reflection to perform human activities and object sensing. A sleep monitoring system proposed by Nandakumar *et al.* [24] utilizes FMCW acoustic signal reflected from human chest to track the breathing rate of a user. However, the study only focuses on detecting partial human body movements and requires the phone to be placed at a fix position near the sleeper. FingerIO [25] uses OFDM acoustic signal reflected from the fingers to localize 2D finger movements and recognize finger gestures [25]. Although it provides high accuracy for finger tracking, it only works for limited range at about 25cm. Moreover, CovertBand [26] utilizes acoustic signals emitted from the phone to track human body activities behind the wall. It is however designed for sensing the body movements instead of stationary obstacles such as the walls and furniture. AIM [21] proposed by Mao *et al.* provides acoustic imaging of an object by moving a acoustic emitting device around the object. Their system is inapplicable to obstacle detection scenarios as it requires to move the phone with pre-defined trajectory around the object. Although the above works all rely on acoustic signals reflections, they are very limited in obstacle collision detection as they either focus on human movements instead of stationary objects, or work only with small sensing range at centimeter level, or require move the mobile devices with pre-defined trajectory.

In addition, a lot of works have been done focus on direction estimation, due to the direction estimation usually use the time differences of signals arriving on different antennas, these works usually require additional or specialized hardwares such as antenna array and directional antenna [13, 28, 33, 43]. Besides TDoA, Doppler shift is also used to localize moving objects in the existing studies [3, 42]. However, those methods either require pre-knowledge of the distinctive frequency components or only work with microphone array (i.e., over 4 acoustic sensors).

## 3 SYSTEM DESIGN

To detect possible obstacle collisions, we introduce an acoustic sensing approach that leverages the dual-microphone and the advanced audio hardware on smartphones. In this section, we discuss in detailed design challenges, system overview, the beep design, and the methods used to extract fine-grained information of the obstacle for collision detection.

### 3.1 Design Challenges

Intuitively, we are able to sense the surroundings by analyzing the signal reflections from the obstacles. While the intuition is simple, there are significant challenges in extracting fine-grained information of the obstacle with off-the-shelf smartphones.

**Lack of Information for Collision Decision.** Previous work on smartphone-based acoustic sensing can only detect the presence of the obstacle and/or the distance between the obstacle and the user [37]. Such coarse-grained information is insufficient to provide accurate obstacle collusion detection. For example, it lacks of the walking angle information of the user with respect to the obstacle. Moreover, without taking the obstacle size information into consideration, existing approach produces a large number of false alerts that distract users.

**Body Reflection.** During distracted walking, a smartphone user holds the phone in front of his/her body. The body of the user thus also reflects the emitted acoustic signals. Although such signal reflections will not affect the obstacle detection, they significantly reduce the accuracy of the estimated distance between the obstacle and the microphone that is close to user's body. Without precise time-difference-of-arrival (TDoA) of the reflected signals to the two microphones of the smartphone, accurate obstacle collision detection could be difficult.

**Hardware Limitation.** Unlike radar or sonar systems equipped with dedicated ultrasound transmitters and receiver arrays, the off-the-shelf smartphones are limited by their hardware platform and functionalities. For example, the smartphones have limited number of microphones and speakers, since they are primarily designed for voice and music recording and playing. We thus need to carefully design the beep signals and the detection methods to overcome these limitations.

### 3.2 Approach Overview

The key idea underlying our obstacle detection system is to infer fine-grained information of the frontal obstacle based on the signal reflections received at two microphones of the smartphone. As illustrated in Fig. 1, when the user is holding and using his smartphone while walking, the build-in top speaker keeps sending our designed inaudible beep signals and the two microphones of phone keep recording the direct propagated beep signals as well as the signal reflections from the surrounding obstacles. Our system then analyzes the received signals at two microphones to infer the distance between the user to the closest frontal obstacle, the angle between the user's walking path and the obstacle, and the size of the obstacle for obstacle collision detection. The system alerts the user if a possible collision is detected.

Fig. 2 illustrates our system flow, which consists of four major components: *Obstacle Detection*, *Obstacle Angle Estimation*, *Obstacle Size Estimation* and *Collision Detection*. The recorded acoustic signals first pass through the Obstacle Detection component, which identifies the signal reflections from the closest frontal obstacle and estimates the distance between the user and the obstacle. In particular, we first filter out all reflections from the user's body and the ones caused by parallel objects such as the ground. Then, we focus on the signal reflections corresponding to the closest frontal obstacle. We estimate the distance from the user to the obstacle based on the round trip time of the corresponding reflected signals.

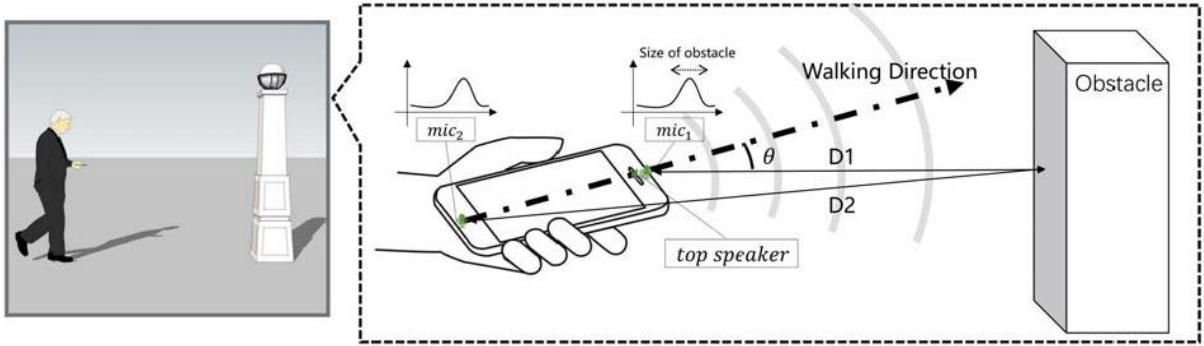

Fig. 1. Obstacle detection using smartphone.

After locating the frontal obstacle, the Obstacle Angle Estimation component analyzes the corresponding signal reflections to infer the angle between the user's walking path and the obstacle. Specifically, it calculates the time-difference-of-arrival (TDoA) of the reflected signals to the two microphones to estimate the angle information. To improve the accuracy of the angle estimation, we apply a multipath mitigation scheme to remove the impact of the user body's reflections, which are mixed with the obstacle's reflections at the microphone.

In the Obstacle Size Estimation component, we leverage the fact that the size of the object has direct impact on the signal reflections to infer the size of the obstacle. Particularly, we utilize the widths (in time domain) of signal reflections corresponding to the frontal obstacle to estimate its size. Finally, our system performs collision detection based on the inputs of obstacle distance, size, and angle.

Our system provides a collision decision and an alert based on a sequence of detections when the distracted user moving forward. In particular, for each beep signal, we perform one collision detection. Our system makes a final decision based on the majority voting of a sequence of detections from multiple beep signals while the user is walking. In addition, we set the dangerous distance to be 3 meters. The system stays silent and keeps running in background if no possible collision is discovered, otherwise an alert will arise if a possible collision is within dangerous distance (i.e., 3 meters).

### 3.3 Beep Design

In designing the beep sound played through the phone speaker, we primarily consider three factors: beep frequency band, length, and unobtrusiveness.

**Frequency Selection.** For object detection systems using radar and sonar technology, the beep signals with wide frequency bandwidths are adopted to enhance the resolution of range estimation [6]. Usually, the wider the bandwidths are, the higher the resolution and accuracy could be achieved. By contrast, the smartphone-based acoustic sensing systems suffer several limitations. First, smartphones have different design purposes other than sonar systems, and they mainly focus on low frequency range to support human voice communication and music. Therefore, the available bandwidths are quite limited. According to our experiments shown in Fig. 3, the frequency response of most smartphones are up to 24kHz, and start to decay quickly after 22-23kHz. Although we could choose up to 24kHz to maximize our frequency bandwidths, the frequency response after 23kHz is weak and unreliable. Therefore, we choose the upper bound of our beep signal as 23kHz. Another factor we consider is that our system is designed to run in background on the smartphone and only alerts the user when a dangerous collision is predicted. Hence for better user experience, the sound signals should be inaudible to human ears so as to minimize the user distraction. The commonly stated range of human hearing is 20Hz to 20kHz, and the upper bound decreases rapidly after 15kHz for older adults. Extending our beep signal to low frequency range could provide us wider frequency bandwidths and better frequency response. However, it will also increases the annoyance. Besides, the frequency of our signal should be kept apart from the environmental noises such as human voices and background noises. Therefore, the lower bound of our beep signal is selected at 18kHz, which enable our beep signal separable from the environmental noise and inaudible to human ears. In summary, given the tradeoff of all the above factors, our system primarily adopts a 18kHz-23kHz bandwidth beep signal. Moreover, we design our beep signal as a linear chirp signal, in which the frequency increases with time. We use a linear chirp signal as it has good autocorrelation property, which enables accurate signal detection under ambient noises.

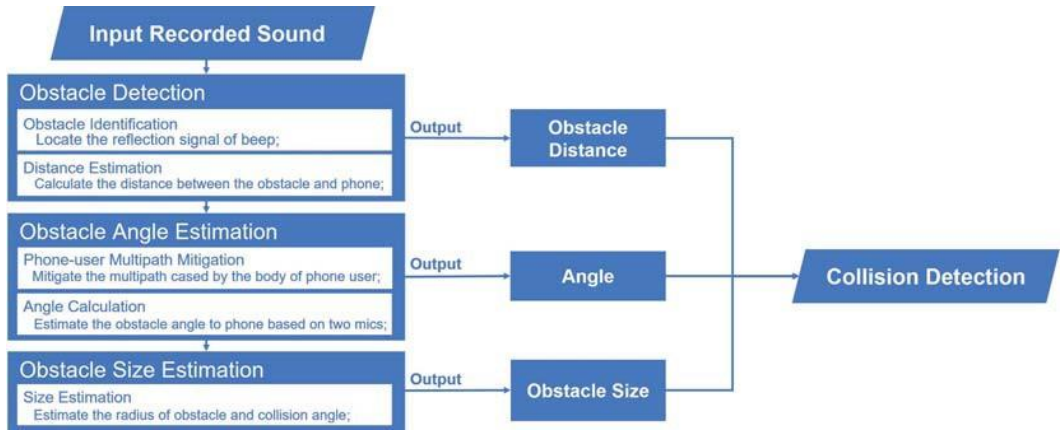

Fig. 2. System Flow.

**Length.** The length of the beep signal impacts both the detection time as well as the quality and reliability of the recorded beep and its reflections. The microphone cannot pick up too short beep signals and their reflections. However, the system detection delay and the multipath distortions could be severe if the beep signal is too long. We found empirically that a beep length of 50ms (9600 samples under 192kHz sampling rate) represents a good tradeoff.

**Unobtrusiveness.** Finally, we apply a Hanning window on our beep signal. It works as a fade in and out window to cancel the the mechanical noises caused by huge frequency changes of the sound wave. Between two consecutive beep signals, an appropriate interval has been planted to avoid the overlap between the beep signal reflections from obstacles. Based on our experimental observations, obstacles that are 7m away have very week reflections. Therefore we design the interval to be 7×2/343m/s = 40ms in our experimental setup.

### 3.4 Obstacle Detection

This step aims to first identify the obstacles. It then calculates the distance between the user and the closest frontal obstacle. Let's denote the top microphone and the bottom one as $mic_1$ and $mic_2$, respectively. The basic idea of identifying the obstacles is to evaluate the correlation between the received signals at $mic_1$ and the original beep signal. Given the received signals at $mic_k$ as $R_k(t)$, where $k=\{1,2\}$ represents either the top or bottom microphone of smartphone, we first apply a bandpass filter to remove the environmental noises and to extract the signal component (denote as $R^{\times}_k(t)$) falls into the beep frequency range (i.e., 18kHz - 23kHz). Then, we calculate the correlation between the filtered signal $R^\times_k(t)$ and the original beep signal with a moving window. The moving window size equals to the length of the beep signal and the correlation is calculated based on the matched filter, a commonly used signal processing technique in radar and sonar system [15]. Basically, we convolve the filtered signal $R^\times_k(t)$ with the transmitted signal, in which

$$C_k(t) = \int_{t^\times}^{t_r} S^*(t^\times) R^\times_k(t+t^\times) dt^\times, \qquad (1)$$

where $S^*(t^\times)$ is the conjugated and time-reversed transmitted beep. We next extract the upper envelop of the resulted correlation (along time series $t$).

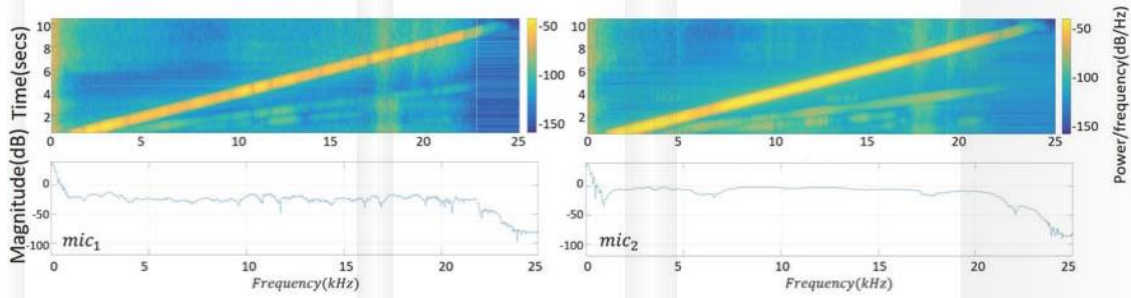

Fig. 3. Frequency response of $mic_1$ and $mic_2$ on smartphone.

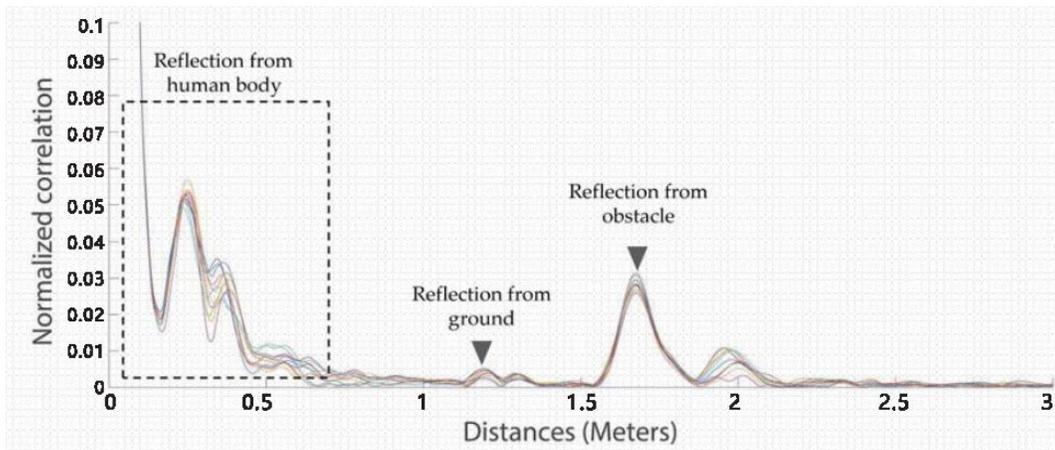

Fig. 4. Obstacle identification.

Fig. 4 shows such an envelop for $mic_1$ when one user standing at one position with obstacles in front of him. We can observe that each peak in the envelop corresponds to one object that reflects signals. Our next step is to identify which peaks are corresponding to the obstacles through peak filtering. With only one static envelop, it is very hard, if not possible, to differentiate the signal reflections between obstacles and other objects, such as the ground and parallel objects to user's walking trajectory (e.g., wall and ceiling). To resolve this issue, we leverage the time series envelop, in which a sequence of envelops will be generated when the user is moving forward. As shown in Fig. 5, when the user is moving towards the frontal obstacle, the distance between the user and the obstacle will be reduced. However, the distances from the user body and ground (as well as other parallel objects like ceiling) to the smartphone are almost remain the same. Based on such an observation, we filter out these peaks that remain at the same locations (on $x$ or $t$ axis) when a user is moving forward. The rest peaks are then identified as the ones resulted from obstacle reflections.

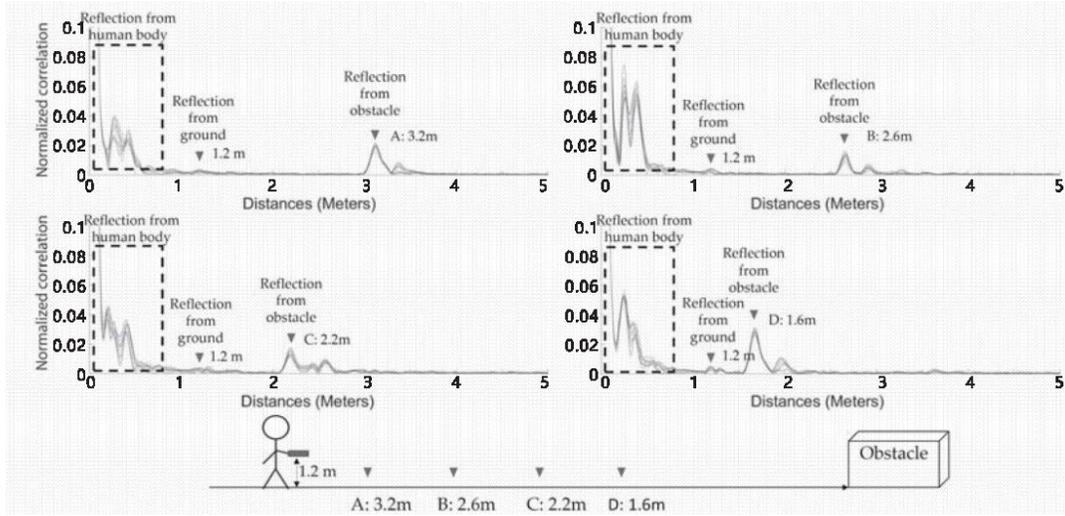

Fig. 5. Moving toward an obstacle.

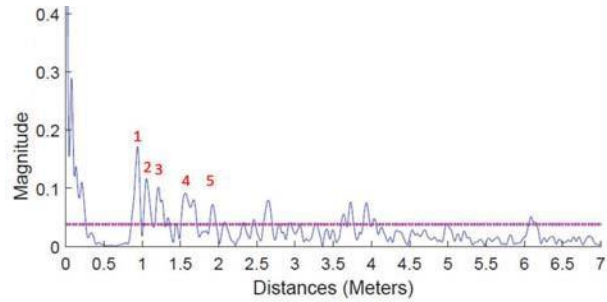

Fig. 6. Multiple obstacle identification.

To identify peaks, we utilize a non-overlapping moving window (i.e. 0.5ms) where the reflection point with the highest magnitude within the window is considered as a candidate peak. Because each round of reflection represents the surrounding environment under current condition, we use the average of correlations within the target range as $threshold_k$. The target range is empirically set to 7m since any reflection outside that range is difficult to discern from noises. To further filter out reflection from insignificant obstacles in the environment, a coefficient $\lambda$ is adopted to adjust threshold in our system. If the candidate peak magnitude is larger than the adjust threshold (i.e., $threshold_k \times \lambda$) calculated from previous steps, we deem such peak represents an obstacle. In addition, it is possible that two peaks of two very close objects (in terms of centimeters, say 10cm) are merged as one peak. In this case, we treat them as one object due to their closeness.

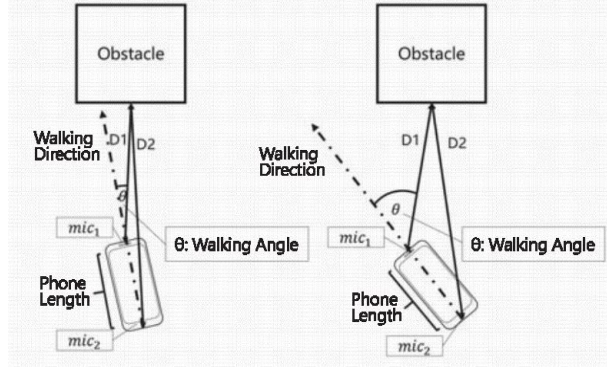

Fig. 7. Estimation of angle between walking path and obstacle.

In realistic walking environment, there may exist multiple obstacles around the user. As shown in Fig. 6, we focus on the top-K closest peaks, which represent the closest obstacles around the user. Certainly, including more peaks could cover more obstacles. However, these far away obstacles are less important and could be covered latter when the user is moving toward them. In our experiments, we choose the top 5 peaks closest to the user. We start from analyzing the closest obstacle and detect the possible collision, and then move forward to analysis the next closest peak. As a result, our system can detect the possible collision with multiple nearby obstacles in time sequence while the user is moving.

After identifying the obstacle candidates, we next calculate the distance between the user and the obstacle by counting the number of delayed samples between the time of sending the beep signal and the time the peak is received. Although we could acquire the time information via system clock or other timer applications, such timing information is unreliable due to various smartphone CPU loads and software delays [44]. Instead, our system count the sample delay between the received direct propagated signal and the received signal reflections from the obstacle. The time point that the signal directly propagates from the phone's speaker to the microphone can be identified as the first strongest peak (i.e. the signal transmits directly from the top speaker to $mic_1$ through the phone body). Considering the top speaker is only several centimeters away from the $mic_1$ and the structure-borne sound propagation speed is much faster than that in the air, we take this time as the beep sending time. Further, the time point of the signal reflection from the obstacle is already identified in the envelop in the last step. Given the number of the delayed samples between these two time point as $m$, the distance $D_1$ (i.e., the distance from the obstacle to $mic_1$) is calculated as $mv/f$, whereas the $v$ is the speed of sound and $f$ is the sampling rate. Assuming the speed of sound is 340m/s, each digital sample represents a distance resolution of 0.18cm and 0.77cm for the sampling rate of 192kHz and 44.1kHz respectively.

### 3.5 Obstacle Angle Estimation

The basic idea of obstacle direction/angle estimation is to leverage the dual-microphone on the smartphone. In particular, the relative angle between the moving direction of the user and the obstacle can be calculated based on the time-difference-of-arrival (TDoA) of the reflected signals to the two microphones of the smartphone. As shown in Fig. 7, given the distances between the obstacle to the two microphones $mic_1$ and $mic_2$ as $D_1$ and $D_2$ respectively, the angle $\vartheta$ can be calculated based on their geometrical relationship as following:

$$\vartheta = \pi - arccos(\frac{D_1^2 + L_{mic}^2 - D_2^2}{2 \times D_1 \times L_{mic}}), \qquad (2)$$

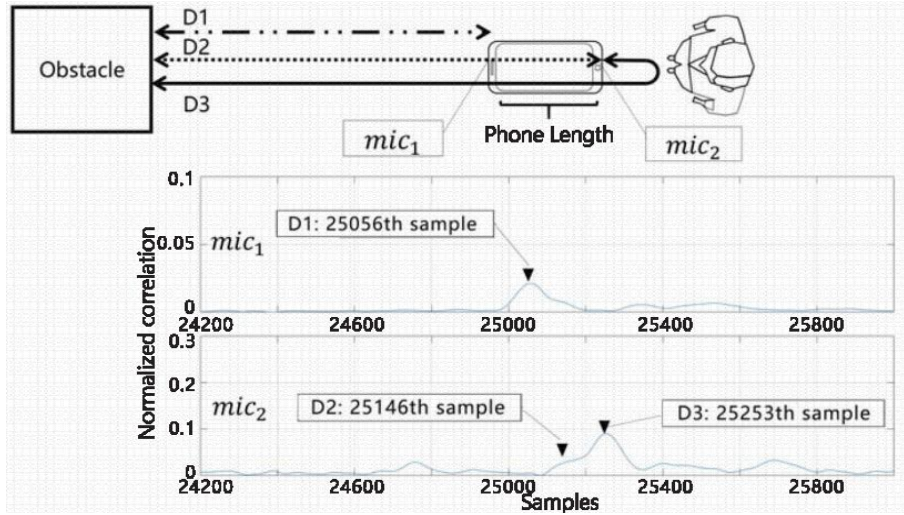

Fig. 8. Multipath effect on $mic_2$.

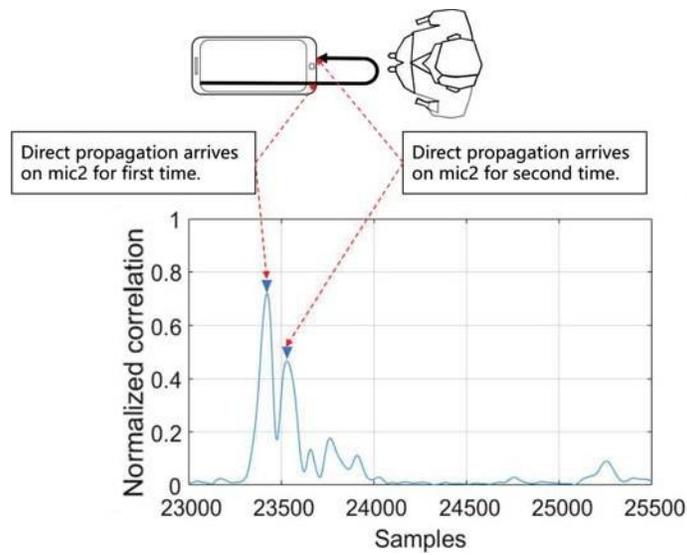

Fig. 9. Multipath mitigation method on $mic_2$.

where $L_{mic}$ is the distance between two microphones on the smartphone.

As we already obtained the distance $D_1$ in the last step, we can use similar way to calculate the $D_2$. However, the estimation of $D_2$ is less accurate compare to $D_1$ due to the following two reasons. First, as shown in Fig. 8, the $D_2$ corresponds to the distance to the bottom microphone, which faces backwards to the reflection direction of the obstacle. Thus, the received reflections from the obstacle are much weaker than these received at $mic_1$. Second, the bottom microphone $mic_2$ is not only facing the user's body but also much closer to the user's body, which leads to strong body reflections combined with the weak obstacle reflections to form one peak (i.e., the peak corresponding to $D_3$ in Fig. 8) instead of two peaks. And such a peak is usually dominated by the strong body reflections (due to the microphone is facing the body). For example, in Fig. 8, we observe that the location corresponds to $D_2$ is not an obvious peak and the actual detected peak corresponds to $D_3$. Thus, the resulted $D_2$ based on the peak detection is actually the $D_3$. Therefore, the body reflections have significant impact on the accuracy of $D_2$ other than the $D_1$.

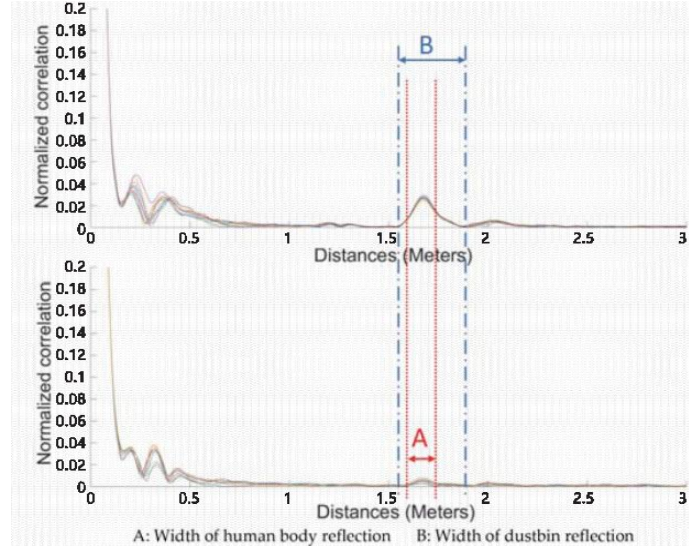

Fig. 10. Reflection of obstacle with different size.

To improve the accuracy of the $D_2$, we need to mitigate the impacts of the user body reflections. We leverage the fact that the $D_3$ corresponds to the obstacle reflections that go through the route between the phone and the user body twice before reaching $mic_2$. Thus, we can deduce the $D_3$ by the distance between the phone and user's body to obtain the $D_2$. To obtain the distance between the bottom microphone $mic_2$ and the user body, we look at the direct propagated beep signal to $mic_2$ and the body reflection based on the direct propagated beep signal, as shown in Fig. 9. Note that there is no obstacle reflections involved in calculating the distance between the bottom microphone and the user body. In Fig. 9, we could observe that the first and strongest peak corresponds to the directly propagated signal (i.e. the beep propagates through the phone's body from the top speaker to $mic_2$) and the close following second peak belongs to the body reflection. Therefore, we estimate the distance between the $mic_2$ and user's body by counting the number of samples between the first and the second peaks as shown in Fig. 9. We then can obtain the distance $D_2$ based on $D_3$ and the the distance between $mic_2$ and the user body. Given $D_1$ and $D_2$, the obstacle direction/angle $\vartheta$ is obtained based on the Equation 2.

### 3.6 Collision Detection

The last piece of information we need for collision detection is the size of the obstacle. To estimate the size of the obstacle, we leverage the fact that the size of the object has direct impacts on the signal reflections. Fig. 10 shows one example of the impact of obstacle size on the width of the reflection crest. At a given distance, we observe that a larger obstacle size (e.g., the dustbin) results in a wider crest width. We experimentally validate such a fact by using obstacles with different size that regularly appear on the road, including human body, signs, dustbin, etc. As shown in the Fig. 11, we calculate the radius based on the width of reflection crest on $mic_1$ and correlate the radius with the ground truth. We find that the correlation is as high as 0.93, which confirms that the object size can be presented by the crest width. Moreover, the distance between the obstacle and the user also affects the crest size. We thus performed another study to investigate the relationship between the crest width and the distance to the obstacle. As shown Fig. 12, the user holds a smartphone and moves toward the obstacle. We observe that as the user gets closer to the obstacle, i.e. distance decreases, the crest width increases. This is consistent with our real life experience, when we are getting closer to the obstacle, the reflection surface are relative larger and we need to make a larger turning angle to avoid collision.

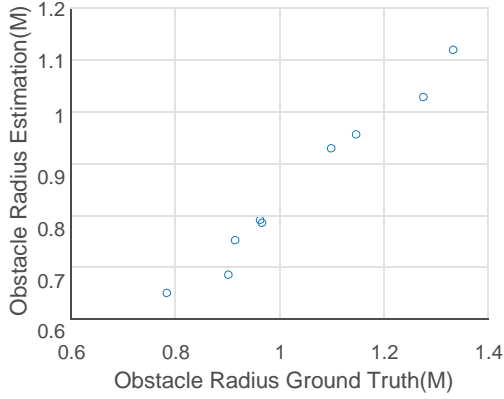
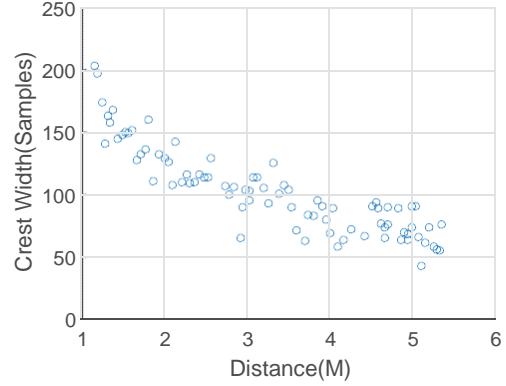

Fig. 11. Correlation between ground truth and estimation of obstacle radius.

Fig. 12. Correlation between distance to user and width of crest.

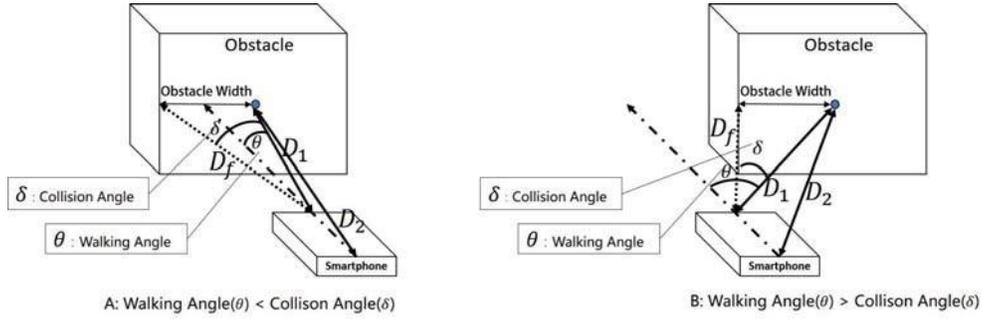

Fig. 13. Walking angle and collision angle under different situation. A: Walking Angle($\vartheta$) < Collision Angle($\delta$) and a collision will happen. B: Walking Angle($\vartheta$) > Collision Angle($\delta$) and a collision will not happen.

We then need to measure the width of the crest that corresponds to the obstacle. To calculate the crest of one peak, we utilize the average of correlation within the target range as $threshold_k$ and count the number of samples within the peak $P$ whose magnitudes are larger than the threshold as crest width $W_p$.

We transfer the width of the crest to infer the size of an obstacle by estimating the collision angle. As shown in Fig. 13, we have the shortest path reflection from the center of the obstacle, as $D_1$. We then could further use the crest width to infer the farthest reflection, i.e. the reflection from the rim of obstacle, as $D_f$, in which

$$D_f = D_1 + \frac{W_p}{2} \times \frac{v}{F_s}, \tag{3}$$

where $v$ is the sound propagation speed in air and $F_s$ is the sampling frequency. Then a collision angle $\delta$ could be calculated using the included angle of $D_f$ and $D_1$, where $\delta = arccos(\frac{D_f}{D_1})$.

After obtaining the distance $D_1$, the walking angle $\vartheta$, and the collision angle $\delta$ which incorporates the obstacle size information, our system performs obstacle collision prediction. In particular, we compare the walking angle $\vartheta$ and the collision angle $\delta$, as shown in Fig. 13, if the walking angle $\vartheta$ is smaller or equals to the collision angle $\delta$, then a collision will happen, otherwise the pedestrian will avoid the collision with the obstacle, that is

$$Decision_{collision} = \begin{cases} true & \text{if } \vartheta <= \delta \\ false & \text{otherwise} \end{cases} \tag{4}$$

When the distance between user and obstacle is within the predefined dangerous distance and decreases gradually, our system will estimate the angle between the user's walking path and obstacle as walking angle $\vartheta$. We then compare this walking angle $\vartheta$ with the collision angle $\delta$ based on the radius of obstacle. If the walking angle $\vartheta$ is smaller or equal to the collision angle $\delta$, a collision decision will be made as true and marked as one collision decision. If the system get five collision decisions out of nine detections in a time sequence, an alert will raise to inform the user, otherwise the user will not walk into the obstacle and it is safe. Our system alerts the user in advance by evaluating the collision possibility when the user is 5 meters away from the obstacle.

## 4 PERFORMANCE EVALUATION

### 4.1 Experimental Setup

**Experiment Environment.** We evaluate our system in two real-world environments including a wide range of obstacle categories.

*Supermarket.* The first place of our experiment is a daily commuting route from a supermarket to an apartment complex with about 500 meters long distance, as shown in Fig. 14. It represents a complex and open environment with different types of obstacles. The obstacle categories include walls, dumpsters, board signs, cars and pedestrians. We show some of these obstacle examples in Fig. 14. They are typical objects with various scales and sizes that may cause injuries to the users when walking distractedly. We conduct experiment following this path 85 times over nine weeks.

*Campus.* The second place we conduct experiment is a campus as shown in Fig. 15. The path we select is a daily route of students and is as long as 2.1km (1.3 miles), which represents a crowded environment. The path includes the following popular locations that students frequently visit: A:parking lot, B:running track, C:student gym, D:library and E:dormitory. There are a number of different obstacles on the path and lots of students walking on the path, especially during the breaks of classes. We conduct experiment following this path 15 times over two weeks.

**Phones and Beeps Selection.** We employ three types of phones including a Samsung Galaxy S5, Samsung Galaxy Note5 and a Galaxy S8+ for the evaluation. We use S5 and S8+ for the scenario of supermarket and S5 and Note5 for the campus. These phones differ in both sizes and audio chipsets. In particular, the length of the phone is 14.1cm for S5, 15.3cm for Note5 and 16.0cm for S8+. The operating systems of these phones are Android 6.0 Marshmallow for S5 and Note5, Android 7.0 Nougat for S8+, which both support stereo recording and playback with up to 192kHz sampling frequencies. We thus choose 192kHz sampling frequency for our evaluation and present the corresponding results. Meanwhile, we experiment with two designed beep signals with 18-23KHz and 16-23KHz bandwidths respectively in the supermarket. Considering the average adult hearing range is 20Hz to 16kHz [5], and we only plays the brief beep signals through the top speaker with low volume, we could assume that our system is transparent to the users.

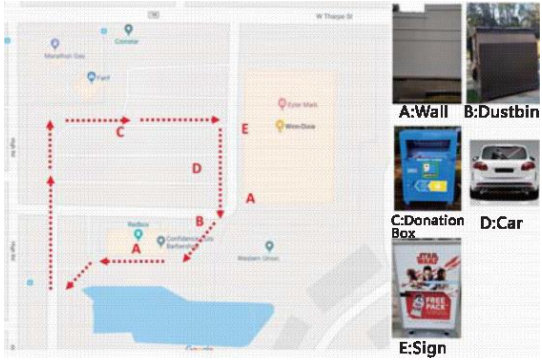

Fig. 14. Experiment path around a supermarket. (A: Walls; B: Dustbins; C: Donation Boxes; D: Cars; E: Signs)

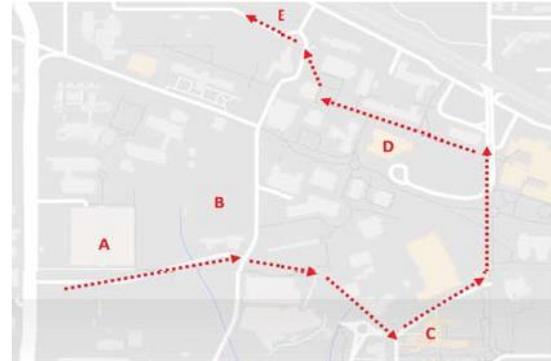

Fig. 15. Experiment path within a campus. (A: Parking Lot; B: Running Track; C: Gym; D: Library; E: Dormitory)

**Data Collection.** Our experiments involve several different categories of objects with various sizes. For each case, the participant walks towards the obstacle and we use camera to record the angle of walking path. We experience 1299 possible cases that collision could happen including 821 cases around the supermarket and 478 cases on the campus. Around the supermarket, we have 108 cases for the walls category, 312 cases of the dustbins, 157 cases of cars, 136 cases of different signs and 108 cases of pedestrians. On the campus, we have 48 cases of walls, 96 cases of dustbins, 106 cases of cars and 228 cases of pedestrians.

**Metrics.** We use the following metrics to evaluate the performance of our collision detection system. True Positive Rate(TP) is the proportion of collisions that are correctly identified and rise alerts. It represents the alert accuracy of the collision detection. True Negative Rate(TN) is the proportion of the non-collisions that don't rise alerts due to the obstacle sizes and user's walking direction.

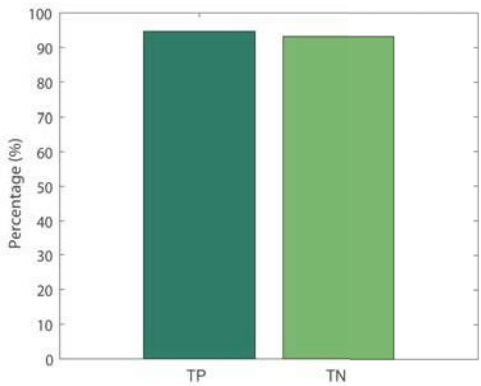

Fig. 16. Overall performance.

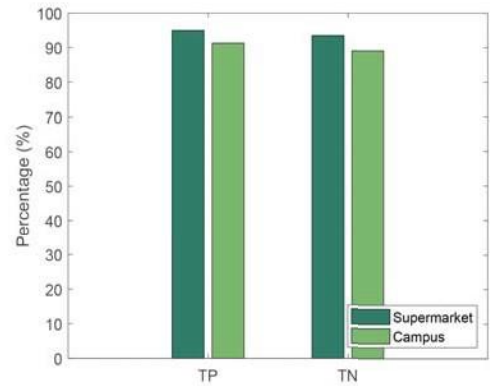

Fig. 17. Accuracy in supermarket and campus.

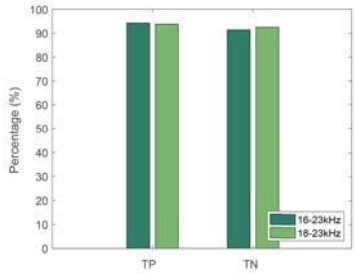 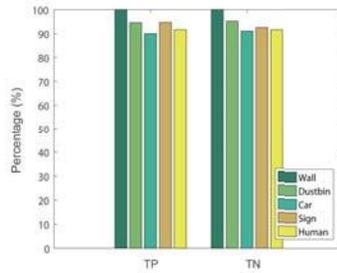 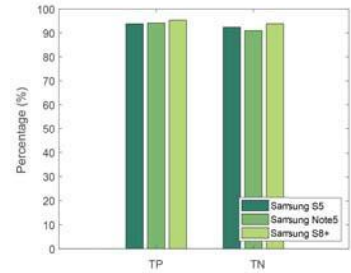

Fig. 18. Accuracy under different frequency band.

Fig. 19. Accuracy facing different obstacles.

Fig. 20. Accuracy using different smartphone models.

### 4.2 Overall Performance

We first evaluate the overall performance of collision detection in both supermarket and campus. Fig. 16 shows the TP and TN when we test our system after we combine results from both supermarket and campus. We observe that the TP of our system, which is the rate of our system to successfully detect the collision and rise an alert, is over 93%. This shows that our system is highly effective and accurate to alert the user of possible collisions if the user keeps the current walking direction. And the TN of our system is over 92% and shows that our system could provide alerts of collisions with low false alerts rates.

Next, we compare the performance of our system in supermarket to that of the campus. The supermarket represents an complex and open space, while the campus environment represent a more crowded environment with many pedestrians. Fig. 17 shows the comparison between TP and TN in these two environments. We observe that the TP and the TN of supermarket is 94.6% and 93.2% respectively. With crowded pedestrians in campus, the TP of our system is slightly reduced to 91.3% and the TN is 89.1%. This is due to the fact that the acoustic signals are affected by a more crowded environment with many other moving pedestrians. Still, our system provides over 90% accuracy in detection possible collisions in a more crowded environment.

### 4.3 Impact of Frequency Band of Beep Signal

Next, under the environment of supermarket, we evaluate the effects of different frequency bands of our beep signals. Specifically, we compare the performance of two different beep signals with 16-23kHz and 18-23kHz frequency bands. As 16-23kHz provides wider frequency band than 18-23kHz, it gives us higher resolution and accuracy in both distance and angle estimations, and results in a better performance in collision decisions. In addition, the extra 2kHz frequency band that 16-23kHz provides has better frequency response on both of our test smartphones, this further contributes to the performance of our system. As Fig. 18 shows, the TP of 16-23kHz frequency band is 94.3%, which is a little bit better than the 93.7% of 18-23kHz. And for TN, 16-23kHz provides 91.4% while 18-23kHz is 92.5%. Overall, both frequency bands could provide satisfactory performance and one could choose either of them depending on his hearing range. To be noticed, we believe that the 16-23kHz beep signal played by the top speaker of the smartphone is imperceptible for most people.

### 4.4 Impact of Obstacles

We further test the performance of our system under different categories of obstacles, these categories are: walls, dustbins, cars, board signs and human, which represent typical obstacles of different sizes that commonly appear in real-world for pedestrians. Fig. 19 shows the comparison results between different sizes of obstacles. We observe that, the larger size the obstacle has, the higher accuracy our system will achieve. The larger obstacles usually have more reflection surface areas, and such obstacles are often built with solid materials, which will produce strong reflections of acoustic signals. One exception is the car, although cars have relative big sizes and are built with metal or other solid materials, the shapes of the cars vary and could affect the performance of our system dramatically. For example, the SUVs usually have larger and flatter reflection areas than sedans, and as a result, our system provides better performance with the SUVs. Another group of obstacles should be noticed is the human beings. Since the reflections of human bodies are evidently weaker than other objects, therefore it is more difficult to capture than that of other categories, hence result in a relative lower accuracy.

### 4.5 Impact of Different Smartphones

As the users of our system may use different phones, we then test the performance of our system using different smartphone models. Specifically, we use S5, Note5 and S8+ to test our system's performance on different types of phones with different sizes, hardwares, and OSs. Fig. 20 shows the TP and TN results under different smartphones. S8+ has longer distance between its two microphones, which contributes to the estimation of included angle between the user's walking path and the obstacle. Also, the frequency response of S8+ on frequency band of our beep signal, is slightly better than S5 and Note5. It turns out the overall performance on S8+ is a little better than S5 and Note5. Even though, our system still produces high accuracy and effective detection results with both phones. Such observations show that our system is robust and compatible to different types of phones.

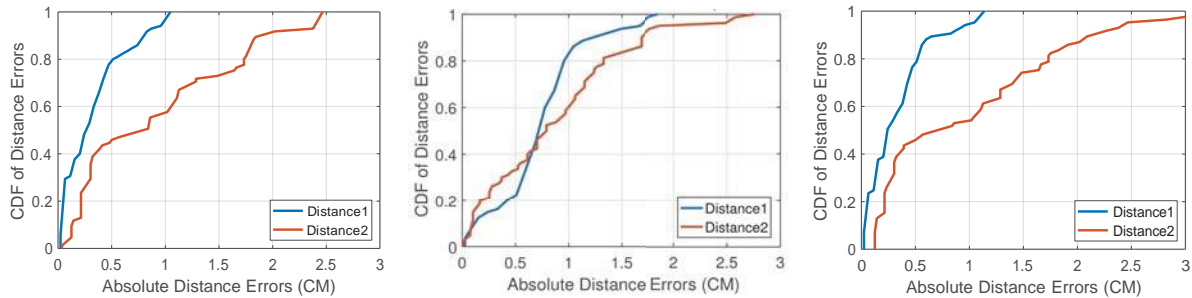

Fig. 21. Distance estimation errors from 1.5m.  Fig. 22. Distance estimation errors from 3m.  Fig. 23. Distance estimation errors from 5m.

### 4.6 Accuracy in Obstacle Detection Under Different Distances

As the distance between the obstacle and the user plays a significant role in obstacle detection, we further evaluate how accuracy our system can detect multiple nearby obstacles with different obstacle distances in terms of obstacle distance estimation. Fig. 21, Fig. 22 and Fig. 23 show the obstacle distance estimation error under three different obstacle distances ranging from 1.5 meters to 5 meters. As we use both top and bottom microphones for distance estimation, we denote the distance from the obstacle to the top microphone $mic_1$ as Distance1 and the distance to the bottom microphone $mic_2$ as Distance2 in the figures. As shown in Fig. 21, our system is highly accurate in estimating the distance between the user and multiple obstacles. Specifically, Distance1 has 90% errors at around 1cm; and Distance2 has 90% errors within 2cm. The distance estimation from $mic_1$ has better accuracy because $mic_1$ always faces the obstacle, and the corresponding recorded signals are dominated by signal reflections from the obstacle. However, the recorded signals at $mic_2$ are significantly affected by user body's reflections.

We next compare the estimation accuracy under different distances. We observe that the distance estimation error increases slightly when increasing the obstacle distance. For example, the median error of Distance1 is

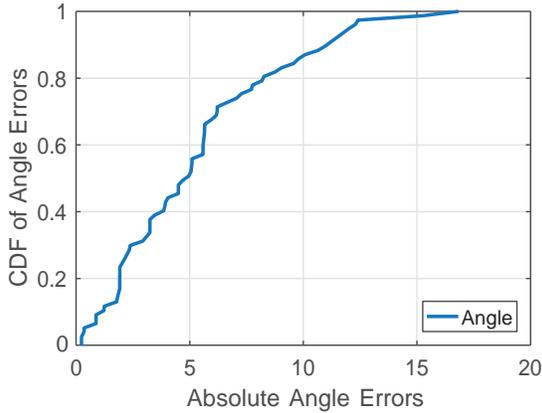
Fig. 24. Angle estimation errors.

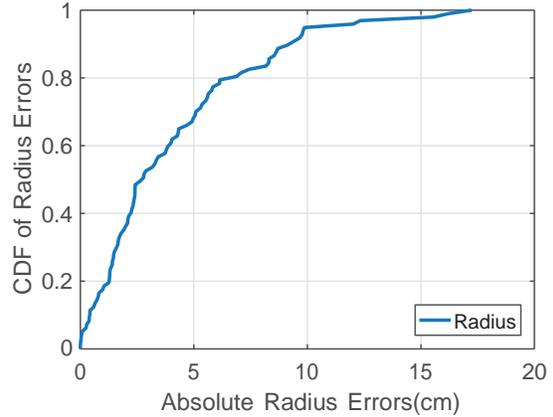
Fig. 25. Radius estimation errors.

increased from 0.25cm to 0.3cm when increasing the obstacle distance from 1.5m to 5m. However, Distance1 is still at around 1cm with 90% probability and the Distance2 is within 2cm with over 90% chance across different distances. This observation is consistent with the collision detection performance, in which we can achieve over 90% accuracy in detecting possible collisions of user within 5 meters away from obstacles .

### 4.7 Accuracy in Angle and Obstacle Radius Estimation

Another important component in our system is obstacle angle estimation, we thus evaluate the accuracy of the angle estimation between user's walking path and the obstacle. To be noticed, we derive the angle utilizing both distance measurements. Thus, the angle estimation result is sensitive to both distance estimation errors. As shown in Fig. 24, our system achieves a median error of 4 degrees with 80% errors within 8 degrees. Such result ensures our system to make accurate collision decisions.

We next evaluate the accuracy for the radius of the obstacle estimation, which would further improve the collision detection accuracy and lower the false alert rate. As shown in Fig. 25, the radius estimation accuracy has a median error of 3cm with 80% errors within 7cm.

### 4.8 Obstacle Detection with and without Fine-grained Information

Our system achieves accurate collision detection by sensing the fine-grained information of the obstacle including the distance from user to obstacle, the angle between the walking path of user and the obstacle and size of the obstacle. We next compare the performance of our system to the one that uses only the obstacle distance and front camera [37]. In particular, existing work detects possible collision when obstacles are detected within 4 meters and there is obstacle captured by the front camera of the phone. Fig. 26 shows the performance comparison between our system that uses fine-grained information and existing work that uses coarse-grained information. We observe that coarse-grained based detection results in large false alert, which is as high as 32.1%. In comparison, our system produces a low false alert which is at around 3.5%. The above result shows that incorporating fine-grained information of the obstacle size and the walking angle can significantly improve collision detection accuracy.

### 4.9 Energy Consumption

Finally, we evaluate the energy consumption of our system. We run our system for 30 minutes on different types of phones (i.e., S5, Note5 and S8+) to evaluate the energy consumption. As shown in Table. 1, on S5, the energy

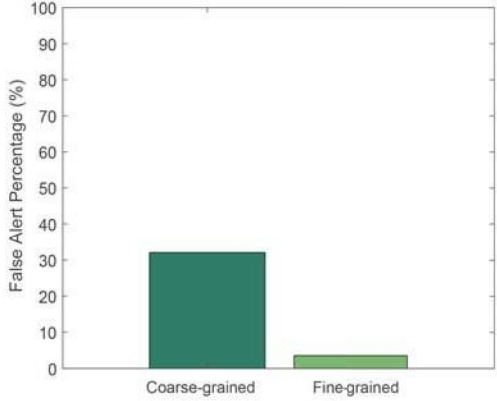 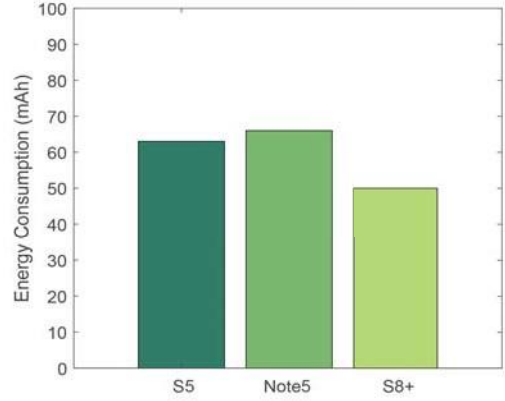

Fig. 26. False alert with coarse-grained information and fine-grained information.

Fig. 27. System energy consumption running on S5, Note5 and S8+ in 30 minutes.

Table 1. Energy Consumption

| Phone | Battery Capacity | Running Time | Energy Consumption | Percentage |
|-------|------------------|--------------|--------------------|------------|
| S5    | 2800mAh          | 30 minutes   | 64mAh              | 2.3%       |
| Note5 | 3000mAh          | 30 minutes   | 66mAh              | 2.2%       |
| S8+   | 3500mAh          | 30 minutes   | 50mAh              | 1.4%       |

Energy consumption of ObstacleWatch running for 30 minutes on different phone models.

consumption is 64mAh, which is about 2.3% of its total battery capacity, and on Note5, it takes 66mAh to run for half an hour , which is about 2.2% of total battery capacity of the Note5. While on S8+, our system costs 50mAh for half an hour running and this takes 1.4% of the phone's battery capacity. As shown in Fig. 27, the energy cost of S8+ is slightly lower than that of S5 and Note5. Results show that energy burden of our system is at users' affordable level, since our system does not run all the time and will only be triggered under the condition when the user is walking and using the phone at the same time. Another factor could be considered is that the energy consumption of the sensors on smartphones will fall if various applications are using the same sensors concurrently [16]. Therefore, when the users are walking with multiple tasks running on their phones, the percentage of the energy consumption of our system will further decrease.

## 5 DISCUSSION
### 5.1 Holding Position of Smartphone
We evaluate our work when user holding the phone vertically, since it is the common holding position adopted by most smartphone users for daily usage. However, there still exists scenarios where the user may hold the phone horizontally (i.e., watch videos, play games). Such holding position will give us the opportunity to identify whether the obstacles are on the left or right side. On the contrary, holding the phone horizontally will reduce the distance estimation accuracy of our system due to microphone facing direction change. Thus, it is important to find the balance between collision detection accuracy and the holding position, which will be included into future works.

## 5.2 Other Potential Threats

In this work, we mainly focus on frontal obstacle collision detection to improve pedestrian safety during distracted walking. Though besides obstacle collision, there exists several other potential threats that might endanger the user. Those threats include: imperfection of road surface (i.e., gap, hole), ramp on the side walk, other pedestrians or vehicles from side or back and so on. Without proper handling, such threats can lead to serious consequences. But they are outside the scope of our work and there are existing work trying to mitigate those threats [11, 12, 34].

## 5.3 Nearby Devices Interference

Our system can achieve good performance when there is no other interferences within the same environment. Though it is possible that there are multiple devices transmitting beep signals concurrently nearby during the obstacle collision detection process. Such interference can be mitigated by decreasing the correlation utilizing multiple access method (i.e., FDMA) where each device transmits its beep signal at a single frequency band. We will also incorporate this into future work to improve our system performance under complex environment scenarios.

## 6 CONCLUSIONS

In this paper, we design a system called ObstacleWatch that leverages the advanced audio hardware of smartphones to sense the surrounding obstacles and infers fine-grained information about the frontal obstacle for collision detection. We find that by sending and recording a well-designed signal beep, the distance and relative position from the phone user to the obstacle, as well as the size of the obstacle could be estimated by leveraging the built-in speaker and microphones of the phone. Benefits from these three outputs, a collision decision could be made and the user can be alerted if there is a collision threat. Our system experimental evaluation demonstrates that a fine-grained object detecting and collision deciding is viable based on built-in sensors on smartphones in real-world environment. And the evaluation results show the ObstacleWatch could achieve over 92% accuracy in detecting obstacle collisions. Results also show that ObstacleWatch is able to work with different phone models with low energy consumption.

## REFERENCES

[1] 2018. New cars are quickly getting self-driving safety features. (Mar 2018). http://saudigazette.com.sa/article/531372/Life/Drive/New-cars-are-quickly-getting-self-driving-safety-features
[2] Corey H Basch, Danna Ethan, Sonali Rajan, and Charles E Basch. 2014. Technology-related distracted walking behaviours in Manhattan's most dangerous intersections. *Injury prevention* (2014), injuryprev–2013.
[3] Yiu-Tong Chan and JJ Towers. 1992. Passive localization from Doppler-shifted frequency measurements. *IEEE Transactions on Signal Processing* 40, 10 (1992), 2594–2598.
[4] Klaus David and Alexander Flach. 2010. Car-2-x and pedestrian safety. *IEEE Vehicular Technology Magazine* 5, 1 (2010), 70–76.
[5] Education.com. 2013. The High Frequency Hearing Test. https://www.scientificamerican.com/article/bring-science-home-high-frequency-hearing/. (2013).
[6] Edward C Farnett and George H Stevens. 1990. Pulse compression radar. *Radar handbook* 2 (1990), 10–1.
[7] Alexander Flach and Klaus David. 2010. Combining radio transmission with filters for pedestrian safety: Experiments and simulations. In *Vehicular Technology Conference Fall (VTC 2010-Fall), 2010 IEEE 72nd*. IEEE, 1–5.
[8] Tarak Gandhi and Mohan Manubhai Trivedi. 2007. Pedestrian protection systems: Issues, survey, and challenges. *IEEE Transactions on intelligent Transportation systems* 8, 3 (2007), 413–430.
[9] Bill Howard. 2014. Ford and Honda stop collisions before they happen with pedestrian detection. (Oct 2014). http://www.extremetech.com/extreme/192863-ford-and-honda-stop-collisions-before-they-happen-with-pedestrian-detection

[10] Tsuyoshi Ishikawa and Kaori Fujinami. 2016. Smartphone-Based Pedestrian's Avoidance Behavior Recognition towards Opportunistic Road Anomaly Detection. *ISPRS International Journal of Geo-Information* 5, 10 (2016), 182.
[11] Shubham Jain, Carlo Borgiattino, Yanzhi Ren, Marco Gruteser, Yingying Chen, and Carla Fabiana Chiasserini. 2015. Lookup: Enabling pedestrian safety services via shoe sensing. In *Proceedings of the 13th Annual International Conference on Mobile Systems, Applications, and Services*. ACM, 257–271.
[12] Shubham Jain and Marco Gruteser. 2017. Recognizing Textures with Mobile Cameras for Pedestrian Safety Applications. *arXiv preprint arXiv:1711.00558* (2017).
[13] Kiran Raj Joshi, Steven Siying Hong, and Sachin Katti. 2013. PinPoint: Localizing Interfering Radios.. In *NSDI*. 241–253.
[14] Keunseo Kim, Hengameh Zabihi, Heeyoung Kim, and Uichin Lee. 2017. TrailSense: A Crowdsensing System for Detecting Risky Mountain Trail Segments with Walking Pattern Analysis. *Proceedings of the ACM on Interactive, Mobile, Wearable and Ubiquitous Technologies* 1, 3 (2017), 65.
[15] John R Klauder, AC Price, Sidney Darlington, and Walter J Albersheim. 1960. The theory and design of chirp radars. *Bell Labs Technical Journal* 39, 4 (1960), 745–808.
[16] Nicholas D Lane, Yohan Chon, Lin Zhou, Yongzhe Zhang, Fan Li, Dongwon Kim, Guanzhong Ding, Feng Zhao, and Hojung Cha. 2013. Piggyback CrowdSensing (PCS): energy efficient crowdsourcing of mobile sensor data by exploiting smartphone app opportunities. In *Proceedings of the 11th ACM Conference on Embedded Networked Sensor Systems*. ACM, 7.
[17] Sugang Li, Xiaoran Fan, Yanyong Zhang, Wade Trappe, Janne Lindqvist, and Richard E Howard. 2017. Auto++: Detecting Cars Using Embedded Microphones in Real-Time. *Proceedings of the ACM on Interactive, Mobile, Wearable and Ubiquitous Technologies* 1, 3 (2017), 70.
[18] Chi-Han Lin, Yi-Ting Chen, Jyun-Jie Chen, Wen-Chan Shih, and Wen-Tsuen Chen. 2016. psafety: A collision prevention system for pedestrians using smartphone. In *Vehicular Technology Conference (VTC-Fall), 2016 IEEE 84th*. IEEE, 1–5.
[19] Jian Liu, Yan Wang, Gorkem Kar, Yingying Chen, Jie Yang, and Marco Gruteser. 2015. Snooping keystrokes with mm-level audio ranging on a single phone. In *Proceedings of the 21st Annual International Conference on Mobile Computing and Networking*. ACM, 142–154.
[20] Xuefeng Liu, Jiannong Cao, Jiaqi Wen, and Shaojie Tang. 2017. Infrasee: An unobtrusive alertness system for pedestrian mobile phone users. *IEEE Transactions on Mobile Computing* 16, 2 (2017), 394–407.
[21] Wenguang Mao, Mei Wang, and Lili Qiu. 2018. AIM: Acoustic Imaging on a Mobile. (2018).
[22] Satish Mohan, Michael E Lockwood, Michael L Kramer, and Douglas L Jones. 2008. Localization of multiple acoustic sources with small arrays using a coherence test. *The Journal of the Acoustical Society of America* 123, 4 (2008), 2136–2147.
[23] Judith Mwakalonge, Saidi Siuhi, and Jamario White. 2015. Distracted walking: examining the extent to pedestrian safety problems. *Journal of traffic and transportation engineering (English edition)* 2, 5 (2015), 327–337.
[24] Rajalakshmi Nandakumar, Shyamnath Gollakota, and Nathaniel Watson. 2015. Contactless sleep apnea detection on smartphones. In *Proceedings of the 13th Annual International Conference on Mobile Systems, Applications, and Services*. ACM, 45–57.
[25] Rajalakshmi Nandakumar, Vikram Iyer, Desney Tan, and Shyamnath Gollakota. 2016. Fingerio: Using active sonar for fine-grained finger tracking. In *Proceedings of the 2016 CHI Conference on Human Factors in Computing Systems*. ACM, 1515–1525.
[26] Rajalakshmi Nandakumar, Alex Takakuwa, Tadayoshi Kohno, and Shyamnath Gollakota. 2017. Covertband: Activity information leakage using music. *Proceedings of the ACM on Interactive, Mobile, Wearable and Ubiquitous Technologies* 1, 3 (2017), 87.
[27] Jack L Nasar and Derek Troyer. 2013. Pedestrian injuries due to mobile phone use in public places. *Accident Analysis & Prevention* 57 (2013), 91–95.
[28] Dragoş Niculescu and Badri Nath. 2004. VOR base stations for indoor 802.11 positioning. In *Proceedings of the 10th annual international conference on Mobile computing and networking*. ACM, 58–69.
[29] Chunyi Peng, Guobin Shen, Yongguang Zhang, Yanlin Li, and Kun Tan. 2007. Beepbeep: a high accuracy acoustic ranging system using cots mobile devices. In *Proceedings of the 5th international conference on Embedded networked sensor systems*. ACM, 1–14.
[30] Z Riaz, DJ Edwards, and A Thorpe. 2006. SightSafety: A hybrid information and communication technology system for reducing vehicle/pedestrian collisions. *Automation in construction* 15, 6 (2006), 719–728.
[31] HC Schau and AZ Robinson. 1987. Passive source localization employing intersecting spherical surfaces from time-of-arrival differences. *IEEE Transactions on Acoustics, Speech, and Signal Processing* 35, 8 (1987), 1223–1225.
[32] Despina Stavrinos, Katherine W Byington, and David C Schwebel. 2011. Distracted walking: cell phones increase injury risk for college pedestrians. *Journal of safety research* 42, 2 (2011), 101–107.
[33] Anand Prabhu Subramanian, Pralhad Deshpande, Jie Gao, and Samir R Das. 2008. Drive-by localization of roadside WiFi networks. In *INFOCOM 2008. The 27th Conference on Computer Communications. IEEE*. IEEE, 718–725.
[34] Maozhi Tang, Cam-Tu Nguyen, Xiaoliang Wang, and Sanglu Lu. 2016. An efficient walking safety service for distracted mobile users. In *Mobile Ad Hoc and Sensor Systems (MASS), 2016 IEEE 13th International Conference on*. IEEE, 84–91.
[35] Leah L Thompson, Frederick P Rivara, Rajiv C Ayyagari, and Beth E Ebel. 2013. Impact of social and technological distraction on pedestrian crossing behaviour: an observational study. *Injury prevention* 19, 4 (2013), 232–237.

[36] Martin Tomitsch and Adrian B. Ellison. 2018. Pedestrian safety needs to catch up to technology and put people before cars. (May 2018). http://theconversation.com/pedestrian-safety-needs-to-catch-up-to-technology-and-put-people-before-cars-65225
[37] Yu-Chih Tung and Kang G Shin. 2017. Use of Phone Sensors to Enhance Distracted Pedestrians' Safety. *IEEE Transactions on Mobile Computing* (2017).
[38] Nisha Vinayaga-Sureshkanth, Anindya Maiti, Murtuza Jadliwala, Kirsten Crager, Jibo He, and Heena Rathore. 2017. Towards a Practical Pedestrian Distraction Detection Framework using Wearables. *arXiv preprint arXiv:1710.03755* (2017).
[39] Christian Voigtmann, Sian Lun Lau, and Klaus David. 2012. Evaluation of a collaborative-based filter technique to proactively detect pedestrians at risk. In *Vehicular Technology Conference (VTC Fall), 2012 IEEE*. IEEE, 1–5.
[40] Tianyu Wang, Giuseppe Cardone, Antonio Corradi, Lorenzo Torresani, and Andrew T Campbell. 2012. WalkSafe: a pedestrian safety app for mobile phone users who walk and talk while crossing roads. In *Proceedings of the twelfth workshop on mobile computing systems & applications*. ACM, 5.
[41] Jiaqi Wen, Jiannong Cao, and Xuefeng Liu. 2015. We help you watch your steps: Unobtrusive alertness system for pedestrian mobile phone users. In *Pervasive Computing and Communications (PerCom), 2015 IEEE International Conference on*. IEEE, 105–113.
[42] Slamet Widodo, Tomoo Shiigi, Naoki Hayashi, Hideo Kikuchi, Keigo Yanagida, Yoshiaki Nakatsuchi, Yuichi Ogawa, and Naoshi Kondo. 2013. Moving object localization using sound-based positioning system with doppler shift compensation. *Robotics* 2, 2 (2013), 36–53.
[43] Jie Xiong and Kyle Jamieson. 2013. Arraytrack: a fine-grained indoor location system. Usenix.
[44] Jie Yang, Simon Sidhom, Gayathri Chandrasekaran, Tam Vu, Hongbo Liu, Nicolae Cecan, Yingying Chen, Marco Gruteser, and Richard P Martin. 2011. Detecting driver phone use leveraging car speakers. In *Proceedings of the 17th annual international conference on Mobile computing and networking*. ACM, 97–108.
[45] Wenyi Zhang and Bhaskar D Rao. 2010. A two microphone-based approach for source localization of multiple speech sources. *IEEE Transactions on Audio, Speech, and Language Processing* 18, 8 (2010), 1913–1928.
[46] Zengbin Zhang, Xia Zhou, Weile Zhang, Yuanyang Zhang, Gang Wang, Ben Y Zhao, and Haitao Zheng. 2011. I am the antenna: accurate outdoor AP location using smartphones. In *Proceedings of the 17th annual international conference on Mobile computing and networking*. ACM, 109–120.
[47] Xiaojun Zhu, Qun Li, and Guihai Chen. 2013. APT: Accurate outdoor pedestrian tracking with smartphones. In *INFOCOM, 2013 Proceedings IEEE*. IEEE, 2508–2516.